\documentclass[12pt]{article}

\usepackage{graphicx}% Include figure files

\newcommand{\beq}{\begin{equation}}
\newcommand{\eeq}{\end{equation}}
\newcommand{\beqa}{\begin{eqnarray}}
\newcommand{\eeqa}{\end{eqnarray}}
\newcommand{\beqar}{\begin{eqnarray*}}
\newcommand{\eeqar}{\end{eqnarray*}}

\begin{document}
\thispagestyle{empty}

\vspace{32pt}
\begin{center}

\textbf{\Large Galactic magnetic fields and the large-scale\\
anisotropy at MILAGRO}

\vspace{40pt}

Eduardo Battaner, Joaqu\'\i n Castellano, Manuel Masip
\vspace{12pt}

\textit{Departamento de F{\'\i}sica Te\'orica y del Cosmos}\\
\textit{Universidad de Granada, E-18071 Granada, Spain}\\
\vspace{16pt}
\texttt{battaner@ugr.es, jcastellano@ugr.es, masip@ugr.es}

\end{center}

\vspace{30pt}

\date{\today}% It is always \today, today,

\begin{abstract}

The air-shower observatory Milagro
has detected a large-scale anisotropy of unknown origin
in the flux of TeV cosmic rays. 
We propose that this anisotropy is caused by 
galactic magnetic fields, in particular, that 
it results from the combined effects of the 
regular and the turbulent (fluctuating) magnetic fields
in our vicinity.
Instead of a diffusion equation, we
integrate Boltzmann's equation to show that the 
turbulence may define a {\it preferred}
direction in the cosmic-ray propagation 
that is orthogonal to the local regular magnetic
field. The approximate dipole anisotropy that we 
obtain explains well Milagro's data.

\end{abstract}

%\pacs{12.60.Jv, 11.25.Mj, 95.35.+d}% PACS, the Physics and Astronomy
%\keywords{Galactic magnetic fields, TeV cosmic rays}
                              %Use showkeys class option if keyword
\newpage

\section{Introduction}

High-energy cosmic rays are of great interest in 
astrophysics, as they provide a complementary picture
of the sky. When they are neutral particles (photons or 
neutrinos), they carry direct information from their
source \cite{Weekes:2008dr,voelk2008,Achterberg:2006md}. Gamma rays, 
in particular, have revealed during
the past 30 years a large number of astrophysical sources
(quasars, pulsars, blazars) in our Galaxy and beyond.
In contrast, when they are charged particles (protons, electrons, 
and atomic nuclei)
cosmic rays lose directionality due to interactions with the 
$\mu$G magnetic fields that they face along their 
trajectory \cite{Strong:2007nh}.
In this case, however, they bring important 
information about the environment 
where they have propagated. For example, the simple observation
that Boron is abundant in cosmic rays while rare in solar
system nuclei is a very solid hint that cosmic rays have crossed
around $10$ g/cm$^2$ of interstellar (baryonic) matter 
before they reach the Earth.

A very remarkable feature in the proton and nuclei
fluxes is its isotropy. 
It is thought that cosmic rays of energy below 
$10^{6}$ GeV are mainly produced in supernova explosions,
which are most frequent in the galactic arms. We observe, 
however, that they reach us equally from all directions.
This can only be explained if their trajectories are close 
to the random walk typical of a particle 
in a {\it gas}, and galactic magnetic fields
seem the key ingredient in order to justify this picture.

Galactic magnetic fields have been extensively reviewed in the 
literature \cite{beck2004,beck2005,wiel2005,Han:2009ts,edu2009}.
It is known that there is an average magnetic field 
of order 
\beq
B_{galactic}\approx 3\; \mu {\rm G} 
\eeq
at galactic scales. 
This component is the background to a second component 
of strength
\beq
B_{random}\approx 3 \! - \! 5\;\mu {\rm G}
\eeq
that is regular within cells of 10--100 pc but changes
randomly from cell to cell. These magnetic fields 
have frozen-in field lines and 
are very affected by the compressions 
and expansions of the interstellar medium produced by 
the passage of spiral arm waves.
A 10 TeV cosmic proton 
would move inside a 5 $\mu$G 
field with a gyroradius 
of
\beq
r_g={p\over e B}\approx 2\times 10^{-3}\;{\rm pc}\;,
\eeq
which is much smaller than the typical region of coherence. 
Therefore, this proton 
{\it sees} the superposition of both components as a 
regular magnetic field:
\beq
\vec B_{galactic}+\vec B_{random}=\vec B_{regular}
\equiv \vec B\;.
\eeq
Notice that the determination of 
the galactic field using WMAP data  
\cite{Page:2006hz,Jansson:2009ip,bea} gives  
$\vec B_{galactic}$. In contrast, estimates from Faraday 
rotations of pulsars 
would be sensitive to the same regular $\vec B$ that
affects the cosmic proton. 
According to Han et al. \cite{han1999,Han:2009ts}, 
the local $\vec{B}$ 
should be nearly contained in the galactic 
plane and clockwise as 
seen from the north galactic pole ({\it i.e.,}
following the direction of the disk rotation),  
although with a small 
vertical component or {\it tilt} angle.

At these small scales the 10 TeV 
proton is {\it diffused} by scattering
on random fluctuations in the magnetic field
\beq
\delta B\ll B\;.
\eeq
The interaction is of resonant 
character, so that the particle is predominantly scattered
by those irregularities of the magnetic field of wave
number $k\approx 1/r_g$. 
Estimates from the standard theory
of plasma turbulence \cite{Casse:2001be} indicate that
$\delta B$ falls as a power law 
for larger wave numbers \cite{han1999}, so 
this component is smaller than the regular $B$.

In this paper we argue that
the detailed observation of the TeV cosmic-ray flux obtained
by Milagro \cite{Abdo:2008aw,Abdo:2008kr} may also provide 
valuable information about $\vec B$ and $\delta B$. 
In particular, the analysis of 
over $10^{11}$ air showers 
has produced a map of the sky showing a large-scale anisotropy
(a north galactic deficit) of order $10^{-3}$.
This map, which is consistent with previous observations 
\cite{Aglietta:1996sz,Amenomori:2006bx}, 
remains basically unexplained. 
Abdo et al. \cite{Abdo:2008aw,Abdo:2008kr} have discussed
several possible origins: 

\vskip 0.2truecm
\noindent {\it (i)}
The Compton-Getting (CG) effect \cite{cg}, a dipole anisotropy that arises 
due to the motion of the Solar System around the galactic center 
and through the cosmic ray background. The 
anisotropy observed in Milagro's map, however, 
cannot be fitted by the predicted CG 
dipole. In addition, the CG anisotropy should be energy independent, 
which does not agree with the data neither.

\vskip 0.2truecm
\noindent {\it (ii)}
The heliosphere magnetic field could produce 
anisotropies \cite{naga1998,schl2007}
that can also be ruled out. The Larmor radius $r_g$ sets the 
size of the coherence cells, and for  
10 TeV protons it is around $2\times 10^{-3}$ pc, 
significantly 
larger than the $5\times 10^{-4}$ pc (100 AU) of the heliosphere. 
Moreover, as pointed out in \cite{Abdo:2008aw,Abdo:2008kr},
the anisotropies persist at higher energies (i.e., for larger 
distance scales), supporting the hypothesis that if  
magnetic fields are involved they are extra-heliospheric. 
\vskip 0.2truecm

Here we explore the effect of the local 
(regular and fluctuating) magnetic fields on the propagation of
TeV cosmic rays reaching the Earth. 
Most analyses model cosmic-ray propagation with a diffusion equation 
\cite{Strong:2007nh,ptus2006,schl2007}, 
assuming certain spatial distribution of sources and a 
diffusion tensor often simplified to an isotropic scalar coefficient. 
This provides the flux over an extended region around the 
solar neighborhood. Here we intend a different approach. The diffusion 
equation derives from Boltzmann's equation, which contains {\it more}
information. The solution of Boltzmann's equation in the vicinity
of the Earth gives   
the statistical distribution function $f(\vec r,\vec p, t)$,  
a quantity related to the intensity or surface brightness used in 
astrophysics. $f$ provides 
the number of cosmic rays per unit solid angle, time and surface  
from any given direction, so it can be compared with 
Milagro's data pixel by pixel.

\section{Cosmic-ray distribution function}

We will treat TeV cosmic rays as a 
fluid that microscopically interacts only with
the magnetic fields, and our objective is
to obtain the 
distribution function $f(\vec r,\vec p, t)$
using Boltzmann's equation.
We will take a basic {\it cell} of radius
$r_g$ and will assume that the non-turbulent component
of the fluid 
\begin{figure}
\begin{center}
\includegraphics[width=9.cm]{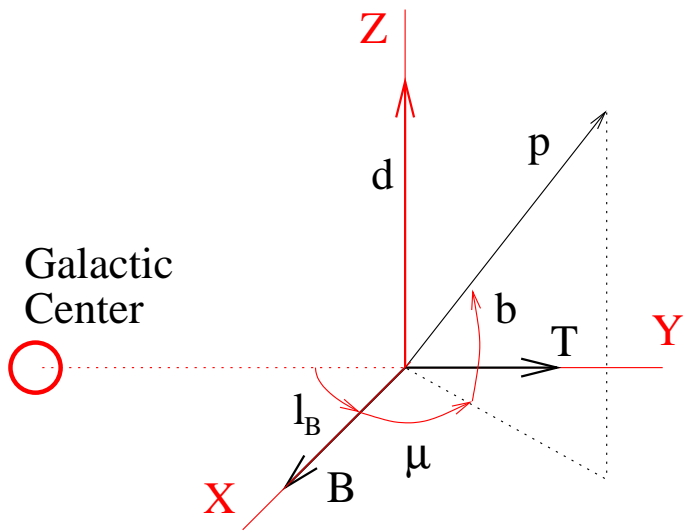}
\end{center}
\caption{Angles $b$ and $\mu$, and orthogonal vectors
$\vec B$, $\vec T$ and $\vec d$ for $l_B=90^o$.}
\end{figure}
is stationary and homogeneous. At these
relatively small distance (and time) scales we 
can also neglect cosmic-ray sources, energy loss, 
or collisions with interstellar matter.
In addition, we take the cosmic rays as
protons (the dominant component in the flux) of 
$E=6$ TeV (the average energy in Milagro's analysis).
Finally, we will assume that the regular 
magnetic field $\vec B$ is on the galactic plane with
a galactic longitude $l_B$, 
although we will show that Milagro's data favors
a component othogonal to this plane (as found in
other observations \cite{han1999}).
In Fig.~1 we have depicted $\vec B$ with $l_B=90^o$.

The frequency of the $(b,\mu)$ direction
in the momentum of cosmic rays reaching the Earth is
then proportional to\footnote{$(E/c)^2 f (\vec u)$ gives
the number of particles with momentum along 
$\vec u$ per unit energy, volume and solid angle
at $E\approx 6$ TeV.}
\beq
f(\vec u)=f(b,\mu)\;,
\eeq
where $\vec u=\vec p/p$, $b$ is the galactic
latitude, and $\mu$ is the longitude relative to the
direction of the magnetic field $\vec B$.
Notice that the galactic longitude of the direction
defined by $\vec u$ is just $l=l_B+\mu$.

Boltzmann's equation 
expresses in differential form
how particles move in the six-dimensional 
phase space \cite{edu}.
In our case this is just
\beq
\vec F \cdot \nabla_{\! u}\; f(\vec u)
= e\; (\vec u \times \vec B) \cdot 
\nabla_{\! u}\; f(\vec u)=0\;.
\eeq
Now, we separate the regular and the turbulent
components both in the distribution function and
the magnetic field:
\beqa
f & \rightarrow & f +\delta f\;,\cr
\vec B & \rightarrow & \vec B +\delta \vec B\;.
\eeqa
The components $\delta \vec B$ and $\delta f$ vary randomly
from one cell to another and have a
vanishing average value,
\beq
\left[ \delta \vec B \right] = \left[ \delta f \right] = 0\;.
\eeq
However, there may be correlations between both fluctuating
quantities. In particular, we will assume a non-zero value of
\beqa
\left[ e\; (\vec u \times \delta \vec B) \cdot 
\nabla_{\! u}\; \delta f \right] &=& 
e\; \vec u \cdot \left[ \delta \vec B\times 
\nabla_{\! u}\; \delta f \right]\cr
&=& e\; \vec u\cdot \vec T
\eeqa
Boltzmann's equation for the regular component is then 
\beq
\left( \vec{u} \times \vec{B} \right) \cdot \nabla_{\! u}\; f +
\vec{u} \cdot \vec T = 0\;.
\label{boltzmann}
\eeq
This equation can be also written 
\beq
\vec{u} \cdot \left( \vec{B} \times \nabla_{\! u}\; f \right)+
\vec{u} \cdot \vec T = 0\;.
\eeq
As $\vec u$ is any direction, this implies 
$\vec{B} \times \nabla_{\! u}\; f = \vec{T}$, {\it i.e.}, the
correlation $\vec T$ must be orthogonal to $\vec B$.
Taking $\vec T$ in the galactic plane, 
\beq
\vec u\cdot \vec T = T\; \cos b\; \sin\mu\;,
\eeq
and expressing 
\beq
\nabla_{\! u}\; f = {\partial f\over \partial b} \;\vec{u}_b + 
 {1\over \cos b}\; {\partial f\over \partial \mu} \;\vec{u}_\mu  
\eeq
with 
\beq
\vec u_b=-\sin b\;\cos\mu \;\vec u_\phi-\sin b\;\sin \mu \;\vec u_r
+\cos b \;\vec u_z\; ; \;\;\;
\vec u_\mu= -\sin \mu \;\vec u_\phi + \cos \mu \;\vec u_r
\;,
\eeq
Eq.~(\ref{boltzmann}) becomes
\beq
-\sin\mu\; {\partial f\over \partial b} +
\tan b\; \cos\mu\; {\partial f\over \partial \mu}
+{T\over B} \; \cos b \; \sin\mu =0\;.
\label{sol0}
\eeq
This equation can be solved 
analytically:
\beq
f(b,\mu)= f_0 \left( 1+{T\over f_0 B}\;\sin b \right)+ 
\tilde f (\cos b\;\cos\mu) \;,
\eeq
with $f_0$ a constant that normalizes $f$ to the
number of particles per unit volume and the second 
term any arbitrary function of the
variable $\cos b\;\cos\mu$.
From the direction $\vec u$ we observe 
cosmic rays with $\vec p=-p\;\vec u$; it is 
straightforward to find the relation between the 
distribution function and the flux $F(b,\mu)$ 
of particles observed at
Milagro per unit area,
time, solid angle and energy:
\beq
F(b,\mu)= {E^2\over c^2} \;f(-b,\mu+\pi)\;.
\eeq
This implies
\beq
F(b,\mu)=F_0 \left(  1-t \;\sin b \right)+ 
\tilde F (\cos b\;\cos\mu) \;, 
\eeq
where $t=T/(f_0 B)$ and $F_0=(E/c)^2f_0$.
Finally, we will expand $\tilde F$ to second
order:
\beq
\tilde F(\cos b\;\cos\mu)\approx 
F_1 \cos b\;\cos\mu + F_2 \left(\cos b\;\cos\mu\right)^2 \;.
\label{sol}
\eeq
The solution in terms
of the galactic longitude is obtained just by
expressing $\mu=l-l_B$.

Several comments are here in order.

{\it (i)} If $F_1=F_2=0$, then the solution
is a dipole anisotropy, 
with the minimum/maximum in the north/south 
galactic poles. This dipole
is then {\it modulated} by the constants $F_{1,2}$, that introduce
an anisotropy proportional to $\cos b\cos\mu$ ({\it i.e.,} the
additional anisotropy coincides along the directions $\vec u$ 
with equal projection on $\vec B$).

{\it (ii)} The dipole anisotropy would vanish if there
were no turbulence ($t=0$): $\vec B$ implies an isotropy broken
by the turbulence in the orthogonal plane.
In contrast, the equation does not say anything about the 
direction along $\vec B$. 
For different 
boundary conditions one can find solutions
with a forward-backward asymmetry (implying diffusion
along $\vec B$) or symmetric solutions. In particular, 
$F_1$ creates an
asymmetry between the $(b=0,\mu=0)$ and 
$(b=0,\mu=180^o)$
directions, whereas the $F_2$ contribution is symmetric.

{\it (iii)} The dominant magnetic field
$\vec B$, the turbulence $\vec T$, and the dipole $\vec d$
are always orthogonal to each other. 
For $\vec B\approx B \vec u_\phi$ the symmetry of the 
galactic disc could favor a radial turbulence, 
$\vec T \approx  T \vec u_r$, like the one that we
have assumed above (see Fig.~1)\footnote{Buoyancy will mainly
produce ascending turbulent cells; since Coriolis forces
are negligible at these small time scales the compression of the 
(frozen-in) azimuthal field lines may result into a $\delta \vec B$
also azimuthal and a vertical $\nabla_{\! u}\; \delta f$, which
imply a radial $\vec T$.}.
However, 
one can change the latitude $b_0$
of the dipole while keeping $\vec B$ on the galactic plane just by
taking the turbulence $\vec T$ out of the 
plane. In particular, the dipole will point towards the
arbitrary direction $b_0$ (see Fig.~2) if 
\begin{figure}
\begin{center}
\includegraphics[width=9.cm]{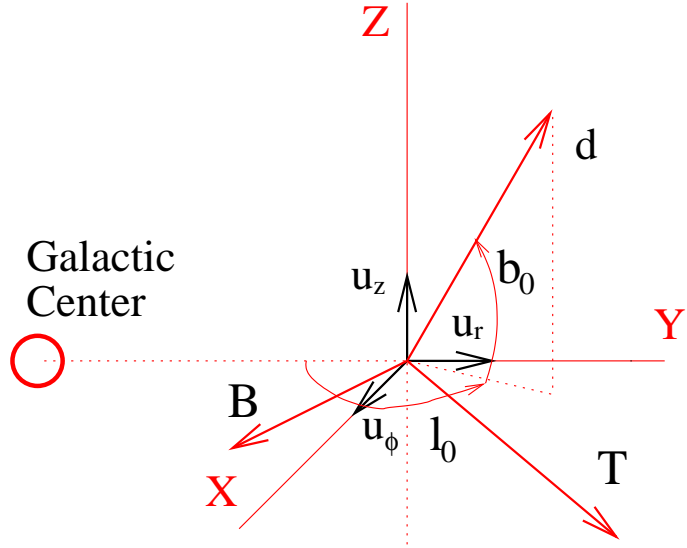}
\end{center}
\caption{Trihedron defined by $\vec B$, $\vec d$ and $\vec T$, 
and coordinate system. 
$\vec B$ is in the galactic plane, whereas $\vec d$ and 
$\vec T$ have latitude $b_0$ and $90^o-b_0$, respectively.}
\end{figure}
\beq
\vec u\cdot \vec T= T \left( \sin b_0 \;\cos b\; \sin\mu-
\cos b_0 \;\sin b \right)\;.
\eeq
The dipole solution is in that case 
\beqa
F(b,\mu) = & F_0 \left[ 1 - t \left(\sin b_0 \;\sin b 
+\cos b_0\; \cos b\; \sin\mu \right) \right] \cr
&+ F_1 \cos b\;\cos\mu + F_2 \left(\cos b\;\cos\mu\right)^2  \;,
\label{sol2}
\eeqa
The galactic latitude $l_0$ of the dipole is then
fixed by the orientation of $\vec B$ in the 
galactic plane,
\beq
l_0=l_B+90^o\;.
\eeq
The direction of the dipole in the basis pictured in 
Fig. 2 is 
\beq
\vec u_d = \cos b_0\; \sin l_0\; \vec u_\phi
-\cos b_0\; \cos l_0\; \vec u_r
+\sin b_0\; \vec u_z
\;.
\eeq

\section{Milagro data}

Milagro data \cite{Abdo:2008aw} indicate 
a clear dipole anisotropy, with
a deficit in the north galactic hemisphere that peaks
at $\delta_0\approx 10^o$ and $AR_0\approx 190^o$ 
({\it i.e.,} $b_0 \approx 72^o$ and 
$l_0 \approx 293^o$). 
In Fig.~3 we plot 
our fit of the data (restricted to a region
in the sky), which is 
\begin{figure}
\begin{center}
\includegraphics[width=13.cm]{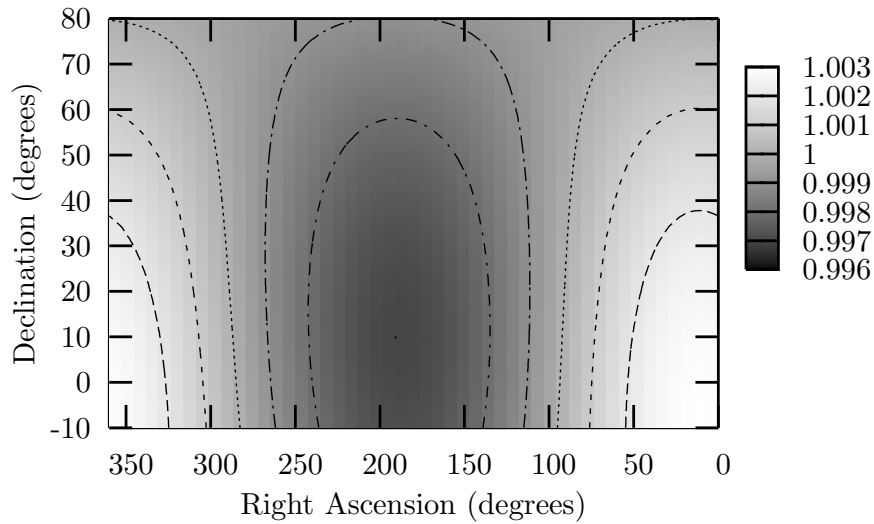}
\end{center}
\caption{
Fit of the Milagro anisotropy. 
}
\end{figure}
obtained for
$t=0.003$ with $F_1/F_0=0$, $F_2/F_0=0.0003$ and a
magnetic field $\vec B$ along $l_B=203^o$.
Our simple fit, an approximate dipole along the direction
of $\nabla_u f$ (from $b_0$, $l_0$ to $-b_0$, 
$l_0-\pi$) provides a good description of 
Milagro's anisotropy.

The fit implies that cosmic rays move near the Earth
with a mean velocity 
\beq
\vec v_0/c= -{1\over N}
\int {\rm d}\Omega \; F(\vec u)\; \vec u
=-0.00059\;\vec u_\phi- 0.00028\;\vec u_r
+0.00157\;\vec u_z
\label{v0}
\eeq
where 
$\;N = \int {\rm d}\Omega \; F(\vec u)\;$ and
the basis is pictured in Fig.~2. 
Eq.~(\ref{v0}) expresses the diffusion velocity
of the fluid (the transport flux $\vec J$ 
is proportional to $N \vec v_0$), and we find that
it goes exactly in 
the direction of the dipole (the term $F_1$ would change 
its direction but we have set it to zero). 

It is important to notice that the regular magnetic field
$\vec B$ does {\it not} need to be on the galactic
plane (our choice above), it can rotate 
around the dipole axis 
and still give the same dipole solution as
far as the turbulence
$\vec T$ is rotated as well. Doing that 
the only changes would appear
in the boundary conditions ($F_1$ and $F_2$), but the 
pure dipole would provide the simplest solution in any case.
The dipole seems to point towards 
\beq
\vec u_d= -0.35 \; \vec u_\phi-0.16  \; \vec u_r+ 0.92 \;\vec u_z\;.
\eeq
Therefore, we can check if this dipole observed at 
Milagro and the local regular magnetic field $\vec B$
(also an observational output) are perpendicular.
We will consider the values of $\vec u_B$
given by Han \cite{han1999,Han:2009ts}. It is found
that $\vec B$ is basically
azimuthal clockwise 
(a {\it pitch} angle of either $0^o$ or $180^o$ 
depending on the definition, which changes for different authors). 
However, the observations 
also indicate the presence of a non null 
{\it tilt} angle of $6^o$ (a vertical component of order 
0.3 $\mu$G) taking 
the magnetic field out of the plane. We obtain an unitary vector 
\beq
\vec u_B= 0.99 \; \vec u_\phi+0.00  \; \vec u_r+ 
0.10 \;\vec u_z\;,
\eeq
which implies a remarkable
\beq
\vec u_d \cdot \vec u_B = -0.18 \;.
\eeq
We think that the approximate orthogonality of 
these two observational vectors (we obtain an angle
of $100^o$) provides 
support to the model presented here.

Notice that our framework could also accommodate 
other anisotropies 
in the flux, added to the dipole one, as far as
they have the same value in all the points with equal projection 
($\cos b\cos \mu$) on $\vec B$.
To explain a {\it pointlike} anisotropy like the
one named as {\it region A} in \cite{Abdo:2008kr}, 
the anisotropy itself 
should be along the direction of the dominant
magnetic field $\vec B$ (orthogonal to $\vec d$). 
{\it Region A}, however, is 
at $(b_A\approx -30^o, l_A\approx 215^o)$, forming
an angle of $58^o$ with the dipole.

\section{Summary and discussion}

Although charged cosmic rays  
do not reveal their source, the study of their
flux from different directions is of interest in astrophysics
because it brings valuable information about the 
interstellar medium. In particular, the {\it per mille}
deficit observed by Milagro could be caused by the
local (at distances of order $r_g$)
magnetic fields.

Using Boltzmann's equation 
we have shown that the interplay between the regular and
the turbulent components in these magnetic fields 
always produces a dipole anisotropy in the cosmic-ray 
flux. 
We find that {\it (i)} the direction of this anisotropy 
is orthogonal to the regular $\vec B$ and {\it (ii)} its 
intensity is proportional to the fluctuations 
$\delta B/B$ at the wave number $k=1/r_g$.
These two simple results have already non-trivial 
consequences. In particular, {\it (i)} implies that
a north-south galactic 
anisotropy would only be consistent with a dominant 
$\vec B$ laying in the galactic plane, whereas 
{\it (ii)} explains that 
the anisotropy is {\it larger} for more
energetic cosmic rays: their gyroradius $r_g$ is 
larger, the resonant wave number $k$ smaller, so 
the expected value of $\delta B/B$ will be larger.

We have argued that Milagro's data can be 
interpreted as a dipole anisotropy
pointing to a well defined direction in the 
north galactic hemisphere,  
namely, $(b_0\approx 67^o,l_0\approx 284^o)$. 
Our model provides a remarkable fit of the data,
so we conclude that it explains satisfactorily 
the large-scale anisotropy found by Milagro.
The model implies that 
the dominant magnetic field near our position 
{\it must} be in the plane orthogonal 
to the dipole ($\vec B$, the turbulence correlation
$\vec T$ and $\vec d$ define a 
trihedron). This plane
forms an angle $\theta=23^o$ with the 
galactic disc. 

The data obtained
by Milagro (energy, direction and nature of 
over $10^{11}$ primaries) shows
that the $10^{-3}$ deficit in the cosmic-ray 
flux from the north galactic hemisphere
already seen in previous 
experiments \cite{Aglietta:1996sz,Amenomori:2006bx}
is actually very close to a dipole anisotropy.
We think that 
the analysis of the flux after substracting
this dipole anisotropy could reveal further
correlations.

\section*{Acknowledgments}
We would like to thank Brenda Dingus for useful discussions.
The work of EB has been funded by MEC of Spain (ESP2004-06870-C02).
The work of MM has
been supported by MEC of Spain (FPA2006-05294) and by Junta de
Andaluc\'\i a (FQM-101 and FQM-437).

\end{document}